# Four Decades of Computing in Subnuclear Physics – from Bubble Chamber to LHC[a]

Jürgen Knobloch/CERN

February 11, 2013

**Abstract**

This manuscript addresses selected aspects of computing for the reconstruction and simulation of particle interactions in subnuclear physics. Based on personal experience with experiments at DESY and at CERN, I cover the evolution of computing hardware and software from the era of track chambers where interactions were recorded on photographic film up to the LHC experiments with their multi-million electronic channels.

## Introduction

Since 1968, when I had the privilege to start contributing to experimental subnuclear physics, technology has evolved at a spectacular pace. Not only in experimental physics but also in most other areas of technology such as communication, entertainment and photography the world has changed from analogue systems to digital electronics. Until the invention of the multi-wire proportional chamber (MWPC) by Georges Charpak in 1968, tracks in particle detectors such as cloud chambers, bubble chambers, spark chambers, and streamer chambers were mostly recorded on photographic film. In many cases more than one camera was used to allow for stereoscopic reconstruction in space and the detectors were placed in a magnetic field to measure the charge and momentum of the particles. With the advent of digital computers, points of the tracks on film in the different views were measured, digitized and fed to software for geometrical reconstruction of the particle trajectories in space. This was followed by a kinematical analysis providing the hypothesis (or the probabilities of several possible hypotheses) of particle type assignment and thereby completely reconstructing the observed physics process.

From the 1970s on, most data were generated (or transformed into) digital form. In Table 1, typical experiments of each decade are listed. The number of scientists involved has increased from a few tens to several thousands. The increasing complexity of detectors required more and more sophisticated software to acquire, reconstruct and analyse the data and to perform the necessary simulations. It is interesting to note that the productivity of software developers stayed rather constant at 10 to 20 lines of code per developer and day.

In this article, I discuss computing aspects of these projects. The final chapters cover the specific topics event display, simulation and data preservation.

---

| Institute | Collaboration | Years | Countries | Institutes | People | Developers | Lines of Code |
|---|---|---|---|---|---|---|---|
| DESY | Streamer Chamber | 1969 - 1976 | 2 | 4 | 20 | 3 | 4 K |
| CERN | CDHS Neutrino | 1976 - 1992 | 4 | 5 | 40 | 4 | 40 k |
| CERN | ALEPH @LEP | 1985 - 2000 | 11 | 30 | 350 | 20 | 600 k |
| CERN | ATLAS @LHC | since 1991 | 38 | 176 | 3000 | 200 | 3-5 M |
| CERN | Grids for LHC | since 2000 | 56 | 340 | 13800 | | |

**Table 1:** Typical experiments and projects of the last four decades

## The Streamer Chamber at DESY

A streamer chamber had replaced the DESY 80 cm bubble chamber because it allowed to trigger on rare events and it allowed linking the obtained pictures of events to external electronic detectors such as a photon beam energy tagging system[1], lead glass shower detectors[2] or proportional chambers[3]. The interactions of high-energy photons and electrons took place in a small hydrogen or deuterium target inside the streamer chamber; therefore the interaction point (vertex) was not visible on the pictures. – see Figure 1.

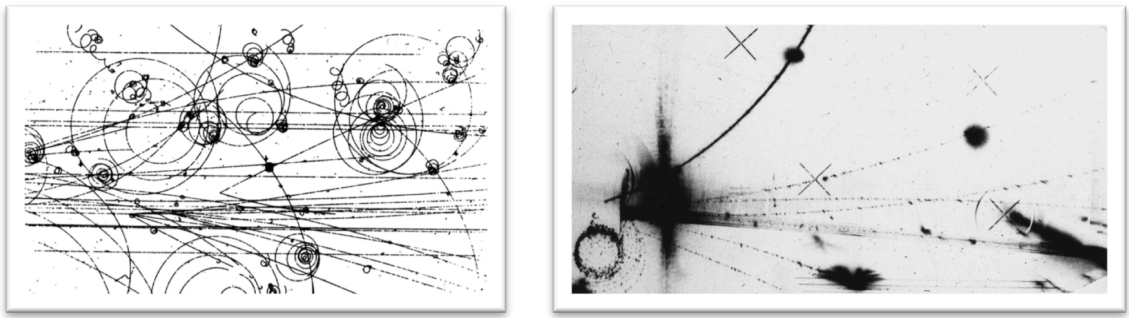

**Figure 1:** Track chamber pictures; left: CERN Bubble Chamber; right: DESY Streamer Chamber

In the early 1970s, the online computing infrastructure for the 20 experiments at the DESY electron synchrotron in Hamburg consisted of on-line computers Digital Equipment PDP-8 connected to the IBM-360 systems in the computer centre. A PDP-8 has an addressing space of 4k 12-bit words. The installations at DESY had four banks of core memory with 4k words each. An innovative operating system[4] developed at DESY provided functionality that became mainstream only years later with the advent of personal computers with windows software and networking. The PDP-8 minicomputers were connected through 24 parallel coaxial cables to a "Parallel Data Adapter (PDA)" in the central IBM 360/75 for data transfer at - for the time remarkable – rate of 100 kbytes/s. Raw data from the experiments were transferred to the mainframe and results of a first analysis in form of histograms sent back to the on-line computer. The PDP-8 was also used to edit programs and to submit them to the mainframe. Documents like doctoral theses could be edited on the PDP and printed on a connected IBM Selectric "golf ball" typewriter.

In the DESY computer centre, a Hierarchical Storage Manager (HSM) provided automatic migration and transparent access to data sets independent of whether they resided on disk or tape.

The beam line was designed with the help an analog computer. From the positions and field strengths of the quadrupole magnets, the horizontal and vertical beam envelopes were determined and could be visualized on a CRT and drawn on a plotter.

The film from the streamer chamber was measured in two different ways. The first one was to project the image on a measuring table. Manually, the operator placed a crosshair on some ten points for each track. These coordinates were digitized for each of the three stereo views and punched on cards to be read on the mainframe computers. The second way employed an automatic device, a flying spot digitizer, the Hough Powell Device (HPD)[5], that could perform many more measurements per track, leading to significantly improved track parameters and the process required much less human intervention.

The measured coordinates were handled by the CERN program THRESH for geometrical reconstruction. The program was modified at DESY to cope with the features of the streamer chamber such as invisible vertex and material distribution. The three stereoscopic views were corrected for optical distortions and demagnification, and corresponding tracks in the views were combined and reconstructed in space. The program GRIND[6] developed by R. Böck at CERN performed kinematical analysis evaluating probabilities for particle mass assignment to tracks.

In the seventies we started to use the most long-living program for physics: the numerical minimization program MINUIT[7] by Fred James. Re-implemented in object-oriented languages it is still used today for a large number of physics studies.

## The Neutrino Experiment WA1 at CERN

Coming to CERN in mid-1976 meant that I had to give up certain habits that were standard at DESY:

- a hierarchical storage management system in the computer centre was still decades away at CERN,
- instead of central data recording, tapes recorded at the experiment were transported to the CERN computer centre – "bicycle on-line"!

In December 1976, the WA1 experiment started data taking in the CERN SPS neutrino beam. The detector consisted of 19 iron modules combining the functions of target, muon absorber, spectrometer magnet and hadron calorimeter interspersed with three-layer drift chambers. The 1250-ton detector was exposed to wide-band and narrow-band neutrino beams up to 400 GeV from 1976 to 1984 studying deep inelastic neutrino scattering on iron (and later also on hydrogen and deuterium). The experiment was a collaboration of CERN, Dortmund, Heidelberg, Saclay and (later) Warsaw, known as CDHS. Physics topics were amongst others: nucleon structure functions, electroweak couplings, QCD, and search for neutrino oscillations. One of the first publications could falsify an effect called "high-y anomaly" where a competing experiment at Fermilab had observed unexpected behaviour in antineutrino events. Two years after the discovery of charmed quarks, neutrino interactions offered a unique possibility to study the coupling of charm to other quark flavours by measuring elements of the Cabibbo–Kobayashi–Maskawa matrix. The most frequent events, charged current interactions, have a negative (leading) muon in the final state for neutrino interactions and a positive one for antineutrinos. Charm manifests itself in an extra muon in the final state mostly of opposite charge to the leading muon. As the magnetic field polarity was set to focus the leading muon, and the second muon generally has lower momentum and was bent away from the axis, the automatic reconstruction initially failed to reconstruct this short track reliably (see an example in Figure 2 – left side). For a first analysis, we reverted to a traditional bubble chamber procedure by measuring the track curvature on printed event displays using curve templates as in the early bubble chamber days. Later, of course, the reconstruction was improved to cope also with the difficult cases. Most computing was done on the CDC 7600 machines at CERN and at Saclay.

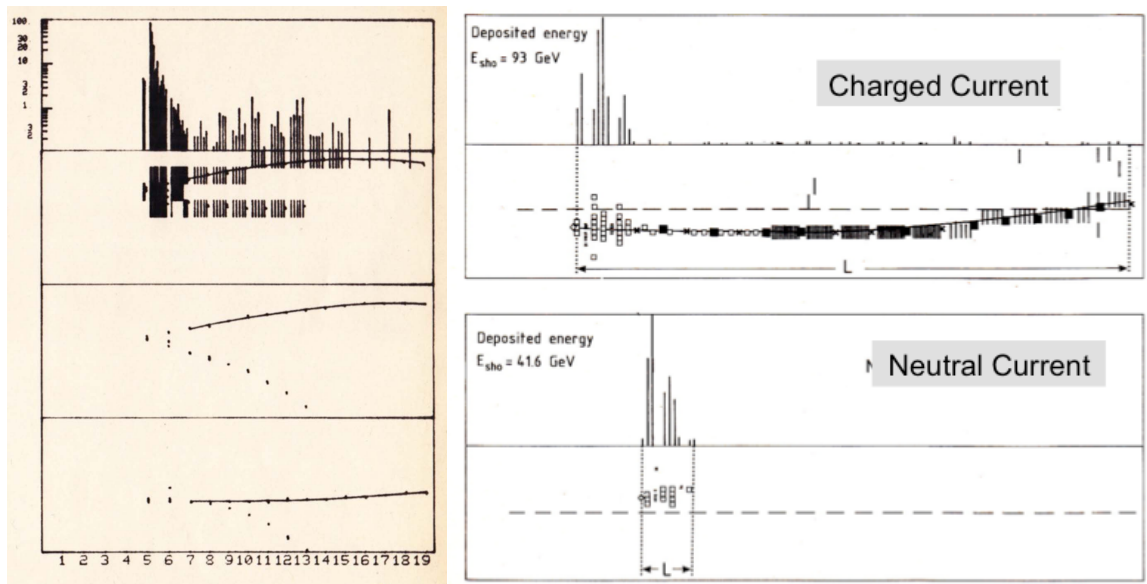

**Figure 2:** Typical neutrino events in the CDHS detector – left: an early di-muon event; top right: a charged current event; bottom right: a neutral current candidate. The events on the right side were recorded in the upgraded CDHSW detector. The histograms give the energy deposited in the scintillators of the calorimeter in units MIP (Minimum Ionizing Particle).

## The ALEPH Experiment at LEP

At colliders such as LEP and LHC, several experiments began operation simultaneously and competed for first results. It was therefore vital that the computing systems (hardware and software) performed to requirements from the first data taking. This was achieved with remarkable success for all experiments at LEP and at LHC. Contrary to the previous experiments with very few sub-detector types, ALEPH[8] had about 10 sub-detectors constructed in the various institutes forming the collaboration. The software was developed in a distributed way between the participating institutes where the partner building the device generally provided the program components specific for that detector. This distributed development required new techniques in the management of software development. In ALEPH, we pioneered in using software engineering techniques[9,10]. External consultants trained the developers in the Structured Analysis, Structured Design (SASD) method. The data were structured according to the entity relationship model as proposed by P. Palazzi for the ADAMO system[11]. The design method allowed decomposing the software projects into components with clearly defined interfaces such that the parts could, in a first step, be developed independently. Though most software was still written in FORTRAN, our system automatically generated code from a formal Data Description Language (DDL) as well as documentation matching exactly the design. The documentation system also generated commented calling trees. Coinciding with the development of the documentation system was the advent of the World Wide Web (WWW) at CERN. This allowed automatic generation of a computing platform independent software documentation with cross-references (hyperlinks) between DDL, tree diagram and code. The system was automatically updated with each new release of the programs for simulation, reconstruction and physics analysis.

With the LEP era, the move from mainframe computers to farms of workstations (UNIX or VMS in the case of LEP, PC in the case of LHC) started to bear fruit. As an interlude, the UA1 experiment had successfully deployed systems of home-made processor boards – emulators. The four LEP experiments initially used different solutions. OPAL started out with a system of APOLLO workstations running UNIX. Their system was subsequently developed into the SHIFT (Scalable Heterogeneous Integrated FaciliTy )[12] system based on SGI and DEC machines in the computer centre. This system was eventually used by all experiments. In ALEPH, we decided to use two distinct operating systems: Digital Equipment VMS running mostly on farms of workstations and IBM/VM on mainframes. Both were used for program development, event processing and simulation. The long lifetime of the experiments at LEP (and even more so now at LHC) exceeds the useful lifetime of a computer platform – operating system – compiler combination. It has proven valuable to avoid the binding to a particular system to later allow moving to a rather different platform. In ALEPH, we could transport our software, documentation and production system rather painlessly to new systems like ULTIX or CRAY[13]. Though the vector performance of the CRAY supercomputer could not really be exploited to full extent for the reconstruction and simulation algorithms, the Cray still acted as key element in the ALEPH computing strategy as fast data server where the vector unit was efficiently used for data compression and conversion.

In ALEPH, we pioneered the reconstruction of events in almost real-time using off-the-shelf commercial processors[14]. A system of VAX workstations processed the events during data taking at the control station of the experiment. This allowed immediate evaluation of calibration and rapid verification of the correct performance of all detectors.

## The ATLAS Experiment at LHC

The detectors at the Large Hadron Collider (LHC) at CERN pose new, extreme challenges for computing:

- the ATLAS detector[15] has 100 million electronic channels;
- bunch crossings every 25 ns delivers 40 MHz of data to the trigger system;
- though the trigger system reduces the rate to a few hundred Hz, the annual recorded data volume is several Petabytes (PB);
- the ATLAS experiment is designed, constructed and operated by a world-wide collaboration of 3000 scientists in 174 institutes in 38 countries,
- 200 software developers have produced about 5 million lines of (mostly C++) code[16].

The computing requirements estimated in 1996 appeared enormous[17,18]; but it was expected that continued validity of Moore's law would ensure that the CPU and storage capacity required would be available about a decade later at LHC turn-on. In fact, the first full year of data taking was only in 2010 when the increase of capacity thanks to technology evolution had taken place such that it could even cope with the new more realistic requirements that exceeded the original plans by two to three orders of magnitude. The increased needs in CPU power and data storage volume were due to better understanding of reconstruction algorithms and to the need for substantial event simulation. Network bandwidth was a considerable concern in the early planning phase. Fortunately, during the LHC construction period, a worldwide privatization and deregulation of telecommunication services led to a massive progress in network performance together with a major cost reduction. Today, contrary to early fears, none of the LHC data needs to be transferred by transportable media such as tapes.

While at previous colliders like LEP, the experiments employed disparate solutions for software and for the computing infrastructure, at LHC an attempt was made to find common solutions. Research and development projects were organized and monitored by the CERN Detector Research and Development Committee (DRDC) and later by the LHC Computing Board (LCB). Only projects approved by at least two LHC collaborations were supported centrally. Some of the projects aiming at "proof of concept" did provide valuable input for later developments. Examples are the RD41 (MOOSE) project aiming to evaluate object-oriented technology and RD45 studying the possibility to use a commercial Object-oriented Data Base Management System (ODBMS) for persistency management. Unfortunately the latter technology – though used for some time in the BaBar experiment at SLAC – could not convincingly be proposed for LHC because it depended on a single vendor who did not find sufficient support in the commercial domain. For decades, during the FORTRAN era, CERLIB provided common tools and packages used by all particle physics experiments and in numerous other sciences. Taking this success to the OO paradigm in the LHC era was the aim of the related LHC++ and CLHEP projects. These projects, though having a rigorous design philosophy were overshadowed by the ROOT[19] project, the object-oriented successor of PAW, which was highly successful in the LEP era. René Brun summarized HEP computing related to analysis tools such as PAW and ROOT in a recent article[20]. The ROOT project benefitted from excellent publicity and from the fact that it chose a subset of C++ as scripting language. Independent but related is the XRootD[21] package from SLAC providing high performance fault tolerant access to data repositories. XRootD is now gaining importance in all LHC collaborations. The most successful project under the LHC R&D developments was Geant4, the simulation framework for LHC discussed in a later chapter.

In experiments prior to LHC, the early data analysis was performed at the site of the accelerator. Only after some time (of about a year), data was also available at outside centres and competitive physics studies could be continued there. For LHC experiments with several thousand scientists it was inconceivable to have all collaboration members come to CERN for the initial running phase. Already the early estimates of the required computing capacity – modest compared to what was actually available at the start of physics running – exceed the possible resources at CERN by a large factor. Early ideas for LHC computing therefore aimed at a hierarchical model with a computing centre at CERN where all raw data were archived and the initial reconstruction was done complemented by several regional centres with significant computing and data storage capacity followed by computing installations at institutes for individual analysis. The groundwork for the topology of the LHC computing infrastructure was laid in the MONARC[22] project. The aim was to determine the requirements and characteristics of the regional centres and to understand the data access pattern during the analysis process. The project developed and implemented simulation tools to understand the characteristics of different implementation models and their principal parameters in terms of data handling, network and CPU capacity. The results of MONARC together with ideas of grid computing as formulated by Foster and Kesselman[23] formed the basis of the LHC computing grid.

## The LHC Computing Grid and European Grids

Two projects leading to the current computing infrastructure for LHC were started in 2001:

- The first in a series of computing grid projects funded through the framework programmes of the European Commission, the European Data Grid (EDG).
- The LHC Computing Grid (LCG) project, based on a proposal by Delfino and Robertson[24], was approved by the CERN Council[25].

The LCG project is a collaboration between the four major LHC physics collaborations, twelve large centres (CERN as "Tier-0" and 11 "Tier-1" centres) and 38 federations of smaller "Tier-2" centres. The task of the Tier-0 at CERN is to provide the raw data recording, a first pass reconstruction and the data distribution to the Tier-1 centres where another copy of the data is permanently stored and where re-processing and physics analysis is done. The 140 Tier-2 sites concentrate on Monte Carlo production and end-user analysis. The basic principles of LCG (called now WLCG to emphasize its Worldwide character) are documented in a Technical Design Report [26]. Apart from forming collaboration between computing sites, LCG ensures the network connections - initially at least 10 Gbit/s from Tier0 to each Tier-1. The grid middleware (the "operating system" of the grid) is developed and maintained jointly with the European and US grid projects. One major achievement is the interoperation of both middleware implementations.

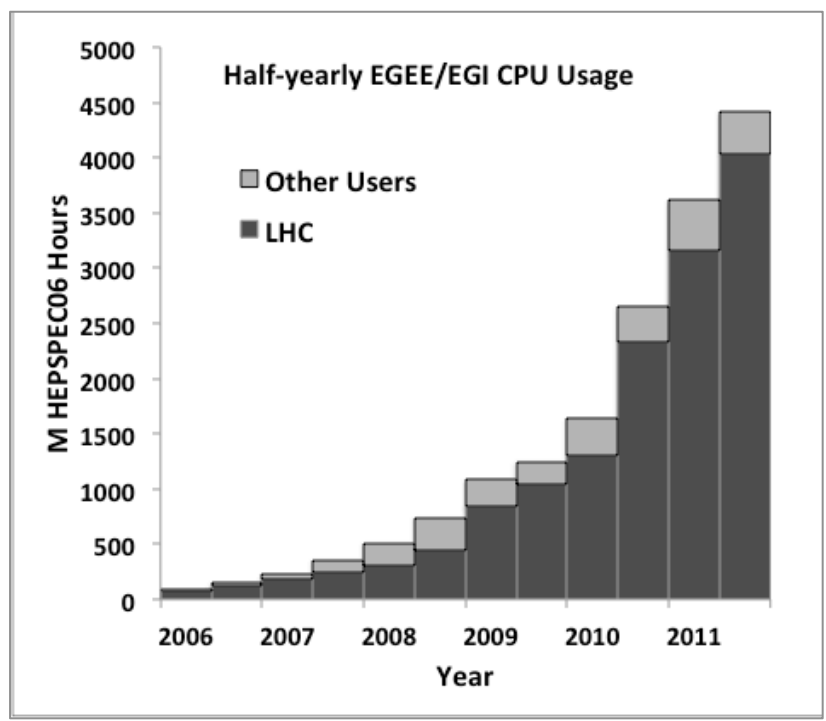

**Figure 3:** Evolution of CPU capacity of European Grid Projects EGEE and EGI

The European grid projects EDG, EGEE, and EGI had a strong CERN and HEP component but also constituted a fruitful partnership of many scientific disciplines including Archaeology, Astronomy & Astrophysics, Civil Protection, Computational Chemistry, Computational Fluid Dynamics, Computer Science, Condensed Matter Physics, Earth Sciences, Fusion, High Energy Physics, and Life Sciences. The rapid evolution of the grid resources provided through these projects is visualized in Figure 3 showing the CPU evolution from 2006 to 2011.

To set the scale, the last column in Figure 3 represents about 100,000 of the currently most powerful CPU cores. It is obvious that the resource usage (and its provision) is dominated by LHC, but e.g. in 2008 life sciences made significant use of the infrastructure for an application to evaluate drugs against the avian flu.

# Event Display

In the era of photographic recording of interactions in track chambers, the film itself represented the visual image of the tracks. Human inspection was essential to understand the physics process behind each event. With the advent of electronic detectors, software reconstruction programs performed the reconstruction. Visualisation of the events was necessary to evaluate the performance (reconstruction efficiency) of the programs and to optimize their performance. A first generation of passive event displays evolved into interactive systems, where, through human interaction, coordinates could be assigned to tracks whose parameters were then determined using the integrated fitting procedure. For the CDHS neutrino experiment, we used terminals storing a vector image on the phosphor of the display screen (TEKTRONIX 4012 and 4014) connected to the interactive CERN FOCUS system on a CDC computer. The UA1 and UA2 experiments certified events constituting the discovery of the W and Z bosons on a 3D display system. Myers and Bettels had developed the Pions graphics system[27] based on an object-oriented database of 3-dimensional coordinates. The data could be visualized on Megatek graphics stations allowing dynamic scaling and rotation giving a 3D impression. H. Drevermann invented another technique for ALEPH [28,29]. A popular projection used in many publications of ALEPH events is the Fish-Eye view visualising in a single picture the sub-millimetre resolution of silicon vertex detectors and the muon chambers extending to 10 m from the interaction point. An example is given in Figure 4.

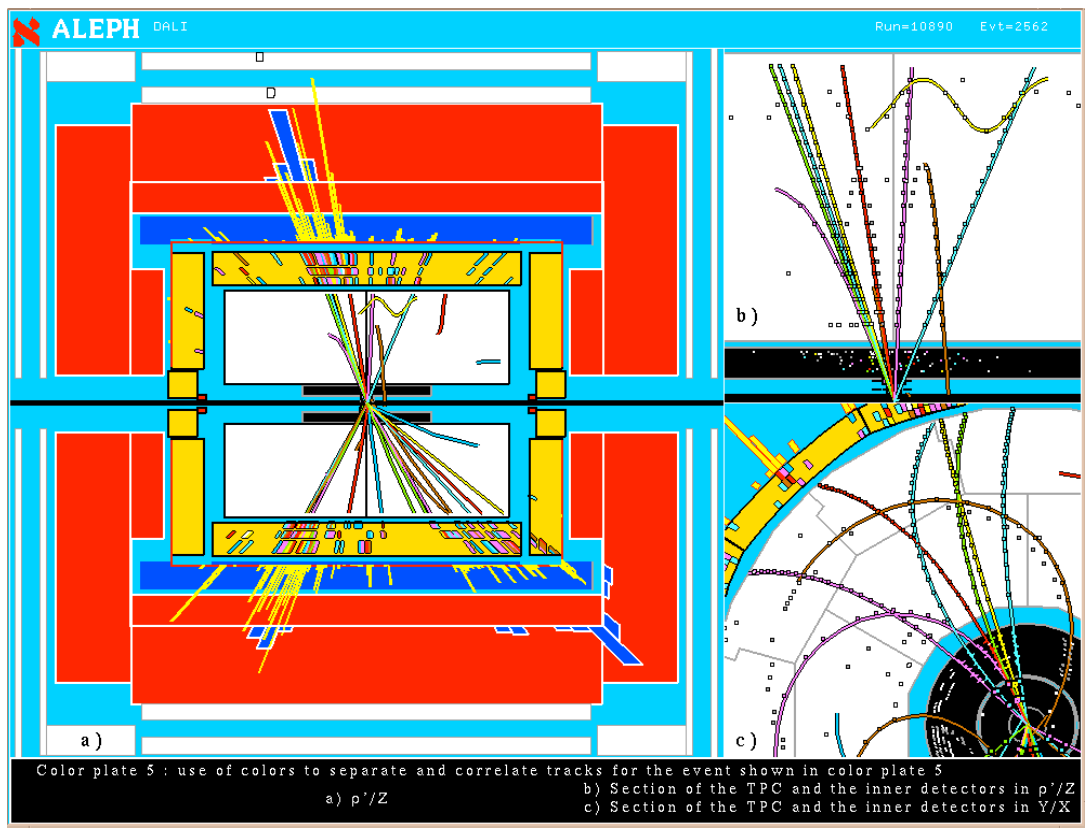

**Figure 4:** DALI event display of an event in ALEPH; the lower right view uses the non-linear radial representation – "fish-eye view"

Another innovative projection is the "v-plot" visualizing, without pattern recognition program, groups of coordinates belonging to individual tracks, indicating whether tracks form a common vertex and giving a measure of the track curvature. These techniques used in the ALEPH event display program DALI are now also implemented in the ATLANTIS event display program[30] for ATLAS shown in Figure 5.

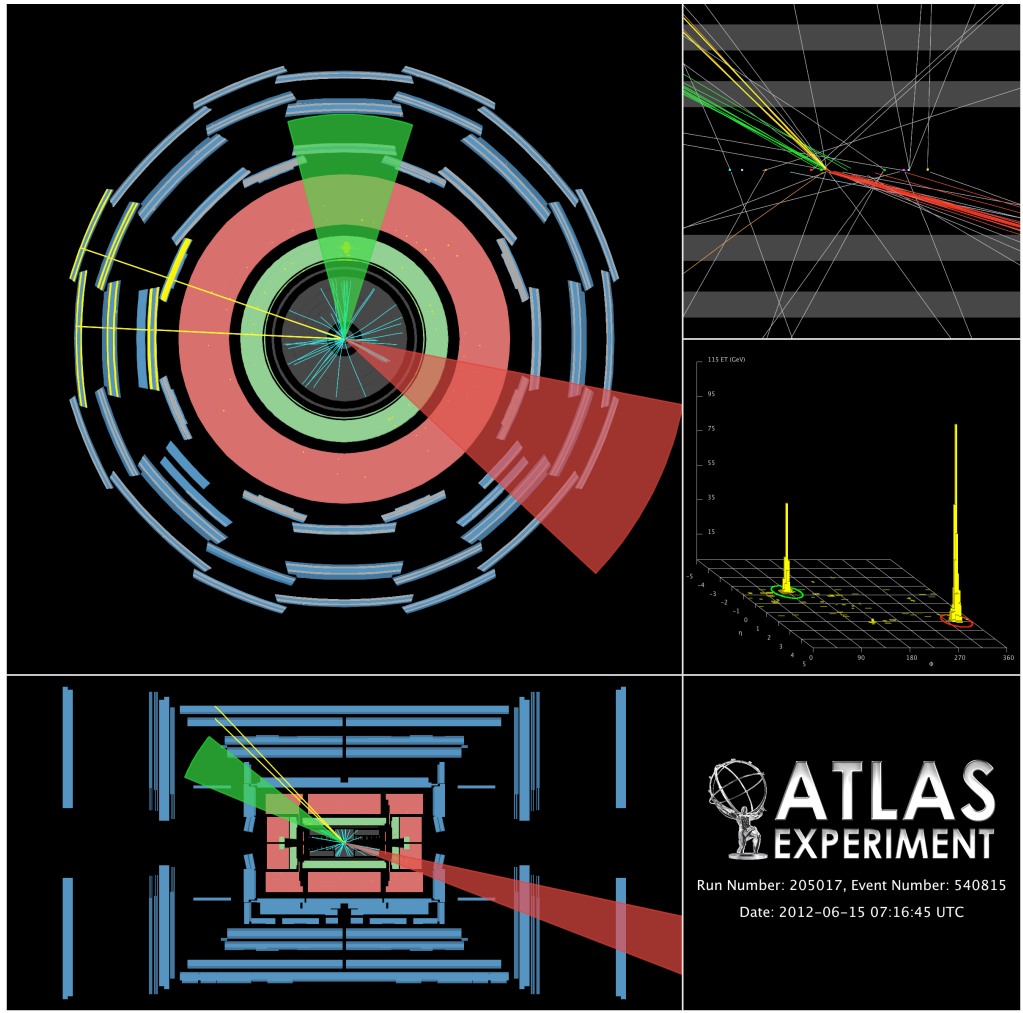

**Figure 5:** ATLANTIS event display showing a Higgs candidate event in ATLAS (ATLAS Experiment © 2012 CERN)

## Simulation

Simulation (Monte Carlo) methods played a major role in nuclear and subnuclear physics even before the advent of electronic computers. An early example was E. Fermi's invention, the FERMIAC[31], a mechanical device to reproduce the stochastic progression of neutrons through various materials. The idea of using computers to simulate particle interactions was first formulated by S. Ulam[32]. During the four decades discussed here numerous simulation programs have been developed. Many of them are based on the groundwork of H. Fesefeldt (GEISHA) for hadronic showers[33] and R. Nelson for Electron Gamma Showers (EGS)[34]. These and other physics codes are now part of simulation packages such as Geant4[35]. Physics generators simulate the primary interactions and feed then into the simulation program, which contains an extremely detailed description of the detectors, giving the distribution of materials and magnetic fields as well as the location and behaviour of the sensitive detector elements. In the simulation process, primary and secondary particles are propagated through the materials and fields of the detectors where interactions and decays are calculated in a stochastic way using pseudo random numbers. The output is formatted such that it can serve as input to the event reconstruction interchangeably with the real data. The simulation programs are used for the design of new detectors, the evaluation of detector acceptance and efficiency and, during analysis, for the estimation of the impact of known background processes on physics observations.

Geant4 was the first large international software project implemented in OO technology following a rigorous design process. It is now the package of choice for the LHC collaborations and of many other subnuclear and nuclear physics experiments. The implemented physics processes are very general and cover a wide energy range such that Geant4 is used for simulations beyond particle physics applications e.g. in the space and medical domains[36]. A recent example is using Geant4 to evaluate the benefits of carbon ion therapy for cancer treatment[37].

The Monte Carlo programs are also used to calculate the induced radiation levels in particle accelerators, experiments and nuclear reactors. This field is dominated by the FLUKA[38] Monte Carlo package with its own physics processes.

## Data Preservation

Experiments in subnuclear physics are global endeavours of the scientific community requiring enormous financial and human resources. As an example, over its lifetime the LHC project will have cost several $10^{10}$ Euros and more than 10,000 scientists will have contributed to construction, operation and exploitation during the about 30 years from project approval to end of data taking. The obtained data and derived results are unique. It is possible that years after the current generation of physicist have terminated the data analysis, new theoretical or experimental evidence needs to be verified or complemented with LHC data. To allow a new analysis of old data it is necessary but not sufficient to keep the data in a format and on media that are still readable. Equally important are the programs for analysis and the metadata describing the detectors and the running conditions. In general, also simulation programs with detailed detector description are required – with the possibility to implement new physics models. In a study I undertook for the 2003 ERPANET/CODATA Workshop on the Selection, Appraisal, and Retention of Digital Scientific Data in Lisbon[39], physicists in several collaborations at CERN and elsewhere agreed that data preservation is important but that any meaningful re-analysis would require the participation of scientists who were members of the original experimental team.

The importance of long-term data preservation is now recognized; it will, however, require significant human and financial resources. A study group DPHEP[40] has been created to evaluate technical and organisational aspects of long-term data usability beyond the lifetime of an active collaboration. The project is supported by physics labs and collaborations and has recently published a status report DPHEP[41] leading to a proposal of a global organisation for data preservation in HEP.

## Summary and Conclusion

We have seen over the past four decades tremendous changes:

The size of physics collaborations has increased from order $\mathcal{O}(10)$ to several thousand. The computing requirements have vastly exceeded the possibilities of a single institution. 40 years ago, one or very few physicists located within earshot did the software development; now, hundreds of developers working on a single package are spread around the globe.

The combined computing capacity of DESY and CERN in 1970 represented one per mill of the power of a single modern smartphone. Over four decades, resources in CPU and storage used for subnuclear physics have increased by eight orders of magnitude. Currently, there are now about $10^{18}$ bits of data storage for the LHC experiments.

To address the main question of this lecture series: "What we would like LHC to give us?" - computing has been essential to enable physics experiments answering the fundamental questions. This has become even more so at LHC than previously. Computing is as important as any other subsystem of the experiments. But also, LHC computing has required pushing technology to new heights enabling other sciences and domains to profit from the new technologies.

## Acknowledgements

In the four decades covered in this article, I had the pleasure to work with wonderful colleagues in an inspiring atmosphere. Naming them all is impossible here – but without them, none of the projects described in this manuscript could have been done.

I greatly appreciate the hospitality of the Ettore Majorana Foundation and Centre for Scientific Culture (EMFCSC). My thanks go to Professor Antonino Zichichi for his kind invitation to give this lecture at the 50th Course of the International School of Subnuclear Physics. Professor Antonino Zichichi and Professor Gerard 't Hooft have led this lively meeting in the splendid city of Erice to a memorable event.